\documentclass[preprint,prd,aps,nofootinbib,amssymb]{revtex4}
\usepackage{graphicx}
\usepackage{epsfig,amssymb,amsmath}

\newcommand{\beq}{\begin{equation}}
\newcommand{\eeq}{\end{equation}}
\newcommand{\bea}{\begin{eqnarray}}
\newcommand{\eea}{\end{eqnarray}}
\newcommand{\bear}{\begin{array}}
\newcommand {\eear}{\end{array}}
\newcommand{\bef}{\begin{figure}}
\newcommand {\eef}{\end{figure}}
\newcommand{\bec}{\begin{center}}
\newcommand {\eec}{\end{center}}
\newcommand{\non}{\nonumber}

\newcommand{\la}{\left\langle}
\newcommand{\ra}{\right\rangle}

\def\GEV#1{10^{#1}{\rm\,GeV}}

\def\lrf#1#2{ \left(\frac{#1}{#2}\right)}
\def\lrfp#1#2#3{ \left(\frac{#1}{#2} \right)^{#3}}
\def\oten#1{ {\mathcal O}(10^{#1})}

\begin{document}
\draft
\tighten
\preprint{TU-945, IPMU13-0176}
\title{\large \bf
Longevity Problem of Sterile Neutrino Dark Matter
}
\author{
   Hiroyuki Ishida\,$^a$\footnote{email: h\textunderscore ishida@tuhep.phys.tohoku.ac.jp},
   Kwang Sik Jeong\,$^a$\footnote{email: ksjeong@tuhep.phys.tohoku.ac.jp}, and
   Fuminobu Takahashi\,$^{a,b}$\footnote{email: fumi@tuhep.phys.tohoku.ac.jp}
    }
\affiliation{
 $^a$ Department of Physics, Tohoku University, Sendai 980-8578, Japan\\
 $^b$ Kavli IPMU, TODIAS, University of Tokyo, Kashiwa 277-8583, Japan
    }

\vspace{2cm}

\begin{abstract}
Sterile neutrino dark matter of mass ${\cal O}(1-10)$\,keV decays into an active neutrino and an X-ray photon,
and the non-observation of the corresponding X-ray line requires the sterile neutrino to
be more long-lived than estimated based on the seesaw formula: {\it the longevity problem}.
We show that, if one or more of the B$-$L Higgs fields are charged under
a flavor symmetry (or discrete R symmetry),
the split mass spectrum for the right-handed neutrinos as well as the required longevity is naturally realized.
We provide several examples in which the predicted the X-ray flux is just below the current bound.
\end{abstract}

\maketitle

\section{Introduction}
One of the central issues in modern cosmology and particle physics is the identity of dark matter.
If dark matter is made of as-yet-unknown species of particles, they must be stable on a cosmological
time scale. The required longevity can be attributed to their light mass and/or extremely weak interactions, and
the elusiveness of dark matter is probably related to its longevity to some extent.
This however  does not necessarily
imply that  dark matter is completely stable; it may have a long but finite lifetime,
decaying into lighter particles. If so, it will enable us to identify dark matter by detecting the signal
of the decay products.

Sterile neutrino is one of the plausible candidates for dark matter, and it has been extensively studied
from various aspects such as the structure formation and baryogenesis.
See Refs.~\cite{Boyarsky:2009ix,Kusenko:2009up,Abazajian:2012ys,Drewes:2013gca,Merle:2013gea} for a review.
Interestingly, sterile neutrino dark matter decays into an active neutrino and an X-ray photon through
mixing with active neutrinos~\cite{Lee:1977tib,Marciano:1977wx,Petcov:1976ff,Pal:1981rm}.
So far, the corresponding X-ray line has not been observed, which places severe constraints on
the mixing angle, or equivalently, its neutrino Yukawa couplings.

The smallness of the neutrino Yukawa
couplings can be partially understood by  a simple Froggatt-Nielsen (FN)
type flavor model~\cite{Froggatt:1978nt} or the split seesaw mechanism~\cite{Kusenko:2010ik},
in which the right-handed neutrinos are charged under a flavor symmetry or propagate in an extra
dimension, while the other standard model (SM) particles are neutral or reside
on the four dimensional brane.
One of the interesting features of these models is that the beauty of the seesaw formula~\cite{seesaw},
which relates the light neutrino masses to the ratio of the electroweak scale to the GUT (or B$-$L) scale,
is preserved even for a split mass spectrum of the right-handed neutrinos, e.g.
$M_1 ={\cal O}(1-10)\,{\rm keV} \ll M_{2,3}$, where $1,2$ and $3$ represent the generation index.
This is because both the light sterile (or right-handed) neutrino mass
and the corresponding neutrino Yukawa couplings are suppressed simultaneously in such a way that the seesaw
formula remains intact.
However, the suppression is not sufficient to avoid the X-ray constraint; the observation requires the
sterile neutrino dark matter to be more long-lived than naively expected. The gap becomes acute for a heavier
mass.
As we shall see shortly, for the sterile neutrino mass of $10$~keV, the corresponding neutrino Yukawa couplings
must be more than two orders of magnitude smaller than estimated based on the seesaw formula.
If there is no correlation among different elements of the neutrino Yukawa matrix as in the neutrino mass anarchy
hypothesis \cite{Hall:1999sn,Haba:2000be}, it would amount to fine-tuning of order $10^{-6}$.
We call this fine-tuning associated with the neutrino Yukawa couplings of the sterile neutrino dark matter
as ``{\it the longevity problem}.''

Taken at a face value, the longevity problem of the sterile neutrino dark matter
suggests an extended structure of the theory, such
as an additional symmetry forbidding the neutrino Yukawa couplings.
In particular, it requires a slight deviation from the seesaw formula for the sterile neutrino dark matter.

In fact, it is well known that, if the sterile neutrino comprises all the dark matter,
its contribution to the light neutrino mass must be negligible in order to satisfy
the X-ray bounds~\cite{Asaka:2005an,Boyarsky:2006jm}.
The point of this paper is to take the observational constraint seriously and construct theoretical models
that could realize both the required split mass hierarchy and the longevity simultaneously.
In Ref.~\cite{Araki:2011zg}, it was shown that the mass spectrum and the mixing angles in the so called
$\nu$MSM~\cite{Asaka:2005an}, where the lightest sterile neutrino has a mass of order keV and
the other two heavy sterile neutrinos have quasi-degenerate masses of ${\cal O}(1)$\,GeV,
can be realized by introducing $Q_6$,
$Z_2$, and $Z_3$ flavor symmetries as well as four SM singlet scalars.
Importantly, the longevity problem was solved in their flavor model.
On the other hand, our purpose is to solve the longevity problem and not to realize the quasi-degenerate
mass for the two heavy sterile neutrinos,
and so, we will consider a relatively simple model
in which the SM is extended by introducing three right-handed neutrinos, a gauged U(1)$_{{\rm B}-{\rm L}}$
symmetry, and an extra flavor symmetry.  Actually one can easily make the lightest sterile neutrino
completely stable by assigning a discrete symmetry such as $Z_2$~\cite{Allison:2012qn},
which however implies that one cannot
observe the sterile neutrino dark matter through its decay.  Also an additional mechanism is required
to realize the split mass spectrum for the sterile neutrinos.
Instead, we will construct models in which a single flavor symmetry realizes both the split mass spectrum
and the longevity of the lightest sterile neutrino. In particular,  the predicted X-ray flux can marginally
satisfy the observational bounds,
so that the X-ray observation still remains a viable probe of the sterile neutrino dark matter scenario.

In this paper we show that the longevity problem can be  solved naturally if one or more of the
B$-$L Higgs fields is charged under a flavor symmetry which also realizes the split mass spectrum, $M_1 \ll M_{2,3}$.
 The main difference from the simple FN model
is that the scalar charged under the flavor symmetry has a non-zero B$-$L charge, and we call such mechanism
achieving the split mass spectrum for the right-handed neutrinos with a sufficiently long lifetime as
``{\it split flavor mechanism}''
in order to distinguish it from the simple FN model.
As we shall see shortly, the split flavor mechanism works well for both continuous
and discrete flavor symmetries, and we provide several examples which solve the longevity problem
and predict the X-ray flux just below the current bound.

\section{Longevity problem}

We consider an extension of the SM with three right-handed neutrinos, and assume the seesaw mechanism~\cite{seesaw}
throughout this paper.
The relevant interactions for the seesaw mechanism are given by
\bea
{\cal L} &=&
i {\bar N}_I  \gamma^\mu \partial_\mu N_I-
\left(
 \lambda_{I \alpha} {\bar N}_I L_\alpha  H
+\frac{1}{2} M_{I} {\bar {N^c_I}} N_I + {\rm h.c.}
\right),
\label{4dL}
\eea
where $N_I$, $L_\alpha$ and $H$ are the right-handed neutrino,
lepton doublet and Higgs scalar, respectively, $I$ denotes the
generation of the right-handed neutrinos, and $\alpha$ runs over the lepton flavor, $e$, $\mu$ and $\tau$.
The sum over repeated indices is understood.
Here we adopt a basis in which the right-handed neutrinos are mass eigenstates, and $M_I$ is set to be real
and positive.
If there is a U(1)$_{{\rm B}-{\rm  L}}$ gauge symmetry, the breaking scale $M$ is tied to the right-handed
neutrino mass, as long as the coupling of the B$-$L Higgs to the right-handed neutrinos is not suppressed.

Integrating out the massive
right-handed neutrinos yields the seesaw formula for the light
neutrino mass:
\bea
\left(m_\nu\right)_{\alpha \beta} = \lambda_{\alpha I}\lambda_{I\beta} \frac{v^2 }{M_{I}},
\label{seesaw}
\eea
where $v \equiv \la H^0\ra \simeq174$\,GeV is the vacuum expectation value (VEV) of the Higgs field.
The solar and atmospheric neutrino oscillation experiments clearly showed that at least two neutrinos
have small but non-zero masses, and the mass splittings are given by  $\Delta m^2_\odot
\simeq 8 \times 10^{-5}\,{\rm eV}^2$ and $\Delta m^2_{\rm atm} \simeq 2.3 \times 10^{-3}\,{\rm eV}^2$.
The seesaw mechanism then suggests that a typical mass scale of the right-handed neutrinos or the B$-$L
breaking scale is around $10^{15}$~GeV, close to the GUT scale, for $\lambda_{I \alpha} \sim 1$.
An attractive feature of the seesaw formula is that it explains the smallness of the neutrino masses
by relating them to the ratio of the electroweak scale to the GUT (or B$-$L) scale.
Furthermore, the baryon asymmetry of the Universe can be generated via leptogenesis by out-of-equilibrium
decays of such heavy right-handed neutrinos~\cite{Fukugita:1986hr}.

The above argument does not necessarily mean that all the right-handed neutrinos have a mass
of order $10^{15}$~GeV.
In fact, it is known that the above mentioned feature of the seesaw formula can be preserved
even for a split mass spectrum of the right-handed neutrinos in the simple FN model~\cite{Froggatt:1978nt}
or the split seesaw mechanism~\cite{Kusenko:2010ik}.
Most importantly, the lightest right-handed neutrino can be dark matter, as it becomes stable
in a cosmological time scale for a sufficiently light mass.
Thus an interesting scenario is that sterile neutrinos have a split mass spectrum $M_1\ll M_{2,3}$ so that
the lightest one contributes to the dark matter while the other two implement leptogenesis.
Intriguingly, this may explain why there are three generations~\cite{Kusenko:2010ik}.

In the simple FN model or the split seesaw mechanism, $N_1$ transforms differently from $N_i$ ($i=2,3$) under
some symmetry or has an exponentially different localization property due to slightly different bulk masses,
respectively.
The mass and Yukawa couplings of the lightest right-handed neutrino $N_1$ are then suppressed as
\bea
M_1 &=& x^2 M,\\
\left|\lambda_{1\alpha}\right| &=& x_\alpha,
\eea
where $x \sim x_\alpha \ll 1$ represents the suppression factor, and $M$ is the U$(1)_{{\rm B}-{\rm L}}$
breaking scale.
The relation $x \sim x_\alpha$ arises from the crucial assumption that the suppression mechanism is independent
of the U$(1)_{{\rm B}-{\rm L}}$ symmetry and its breaking.
The light neutrino masses are still related to the ratio of the electroweak scale to the GUT (or B$-$L) scale,
since the dependence on $x$ and $x_\alpha$ is cancelled in the seesaw formula (\ref{seesaw}) as long as
$x \sim x_\alpha$.

\begin{figure}[t]
\begin{center}
\hspace{-5mm}
\begin{minipage}{8cm}
\includegraphics[width=7.6cm,clip,angle=0]{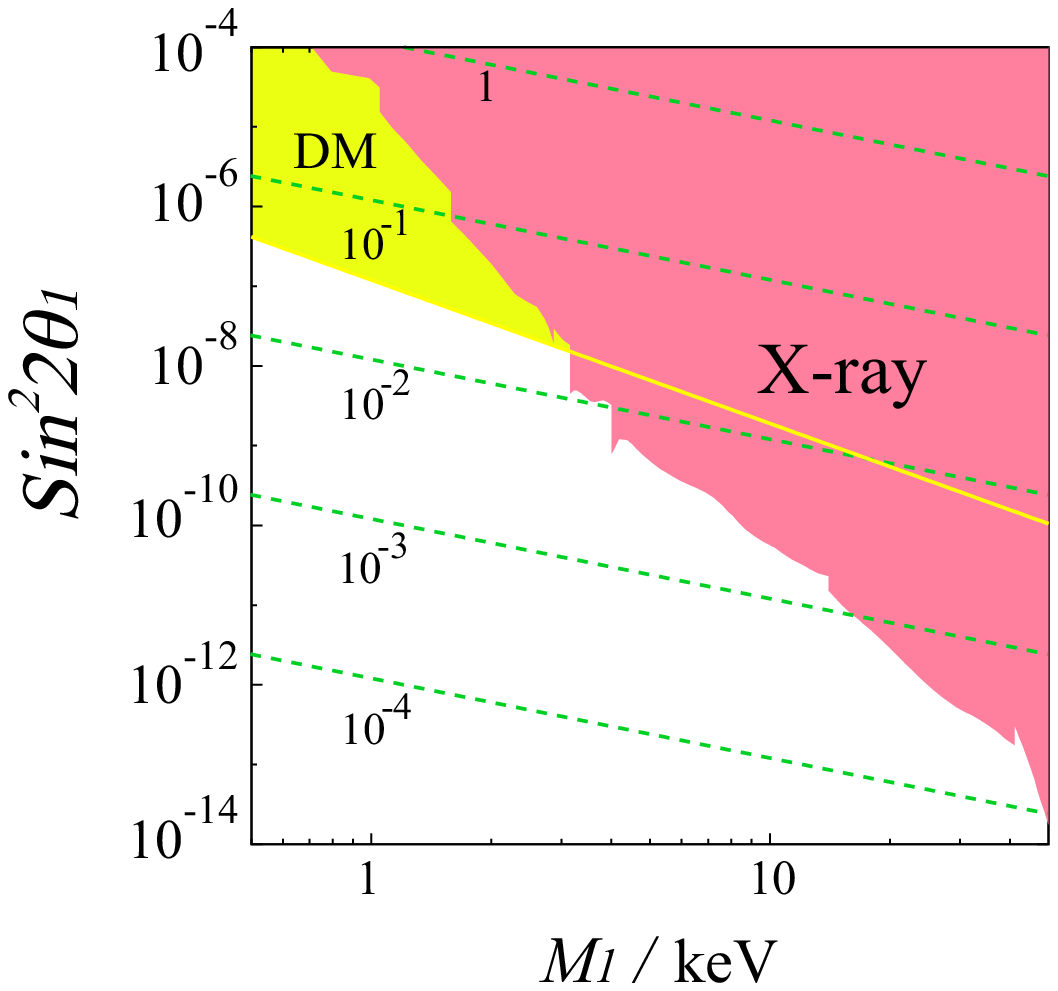}
\end{minipage}
\hspace{-5mm}
\begin{minipage}{8cm}
\includegraphics[width=7cm,clip,angle=0]{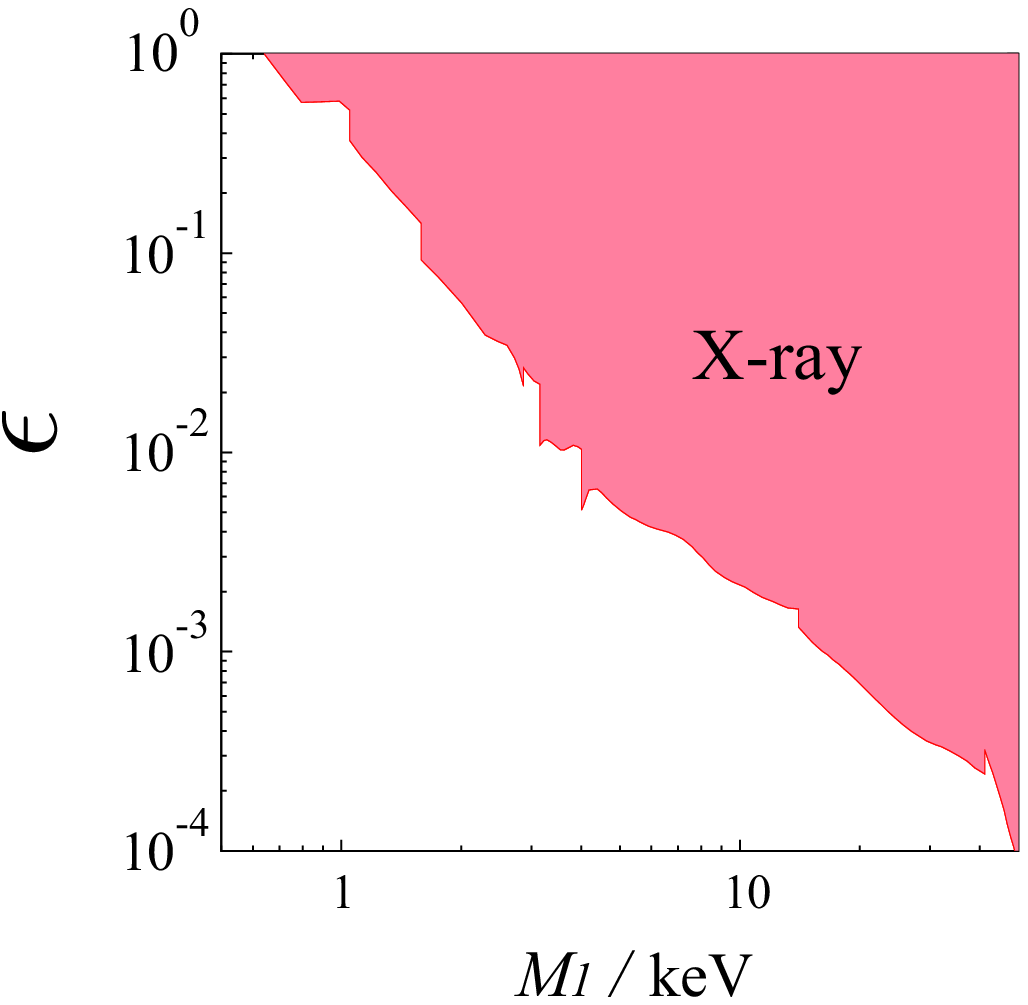}
\end{minipage}
\caption{
X-ray bounds on the mixing angle $\sin^2 2\theta_1$ (left) and $\epsilon$ (right) given as a function
of the sterile neutrino mass $M_1$.
In the left panel, the dashed green lines show the value of $\sin^2 2\theta_1$ estimated by Eq.~(\ref{theta1})
for $\epsilon = 10^{-4},\,10^{-3},\,10^{-2},\,10^{-1}$ and $1$, from bottom to top, respectively.
The upper-right (pink) shaded region in both panels is excluded by the X-ray observations
\cite{Abazajian:2012ys}, while the upper-left (yellow) shaded region in the left panel is excluded by
the dark matter overproduction via the Dodelson-Widrow mechanism \cite{Laine:2008pg,Dodelson:1993je}.
Note that the yellow region becomes viable if there is a late-time entropy production.
}
\label{fig:x-ray}
\end{center}
\end{figure}

On the other hand, the mixing angle between $N_1$ and active neutrinos is given by
\bea
\theta^2_1 &\equiv& \sum_\alpha \frac{|\lambda_{1 \alpha}|^2 v^2}{M_1^2}
\nonumber \\
&=&
10^{-5}\,
\epsilon^2
\left(\frac{m_{\rm seesaw}}{0.1\,{\rm eV}}\right)
\left(\frac{M_1}{10\,{\rm keV}}\right)^{-1},
\label{theta1}
\eea
where we have defined
$\epsilon^2 \equiv \sum_\alpha x_\alpha^2/x^2$, and $m_{\rm seesaw}$ denotes the typical
neutrino mass induced by the seesaw mechanism,
\bea
m_{\rm seesaw} &\equiv& \frac{v^2}{M} \,\simeq\,
0.03\,{\rm eV} \left(\frac{M}{10^{15}\,{\rm GeV}}\right)^{-1}.
\eea
Through the mixing $\theta_1$, the sterile neutrino decays into three active neutrinos,
and also radiatively into active neutrino plus photon~\cite{Lee:1977tib,Marciano:1977wx,Petcov:1976ff,Pal:1981rm}.
The latter process is strongly constrained by the non-observation of the corresponding
X-ray line~\cite{Abazajian:2012ys} (see also Refs.~\cite{Loewenstein:2008yi,Loewenstein:2009cm,Loewenstein:2012px}),
leading to a tight upper bound on the mixing angle as shown Fig.~\ref{fig:x-ray}.
The bound can be conveniently parameterized by~\cite{Boyarsky:2009ix}
\bea
\label{x-ray}
\theta^2_1 &\lesssim& 1.8 \times 10^{-10} \left(\frac{M_1}{10\,{\rm keV}}\right)^{-5}.
\eea
Therefore, $\epsilon$ should be much smaller than unity to satisfy the X-ray bound
for $M_1 \gtrsim$ a few keV:
\bea
\epsilon &\lesssim& 4 \times 10^{-3} \lrfp{m_{\rm seesaw}}{0.1\,{\rm eV}}{-\frac{1}{2}}
\lrfp{M_1}{10\,{\rm keV}}{-2}.
\eea
This requires a deviation from the seesaw formula (\ref{seesaw}) for the sterile neutrino
dark matter $N_1$, and the gap becomes acute for a heavier $M_1$.
Note that the Lyman alpha bounds on $M_1$ reads $M_1 \gtrsim 8$\,keV ($99.7\%\,$C.L.), assuming
the non-resonant production for the sterile neutrino dark matter~\cite{Boyarsky:2008xj}.\footnote{
The bound is relaxed for the production from the singlet Higgs decay~\cite{Kusenko:2006rh,Petraki:2007gq}
or  the resonant production which works in the presence of
large lepton asymmetry~\cite{Boyarsky:2008xj}.
}
Therefore $\epsilon$ must be much smaller than unity, which implies the neutrino Yukawa couplings $\lambda_{1\alpha}$
should be suppressed by about $\epsilon$ with respect to that estimated from the seesaw formula.
For instance, for $M_1=10$~keV, we need $\epsilon$ smaller than $4 \times 10^{-3}$.
If $x_\alpha/x$ takes a value of order unity randomly as in the neutrino mass anarchy, it would require
a fine-tuning of order $\epsilon^3 \sim 10^{-7}$. We call this fine-tuning problem as the longevity problem.
Importantly, the problem cannot be resolved in the split seesaw mechanism or the simple FN model.
As we shall see in the next section, the split mass spectrum as well as the required longevity
can be naturally explained if one or more of the B$-$L Higgs is charged under a flavor symmetry;
the key is to combine the flavor symmetry with the B$-$L symmetry.

\section{Split flavor mechanism}

In this section, we present a modified seesaw model which
realizes the split mass spectrum for $N_I$ while solving the longevity problem.
We consider an extension of the SM with three right-handed neutrinos $N_I=(N_1,N_i)$ for $i=2,3$,
the U(1)$_{{\rm B}-{\rm L}}$ gauge symmetry,
and two B$-$L Higgs fields $\Phi$ and $\Phi^\prime$ whose VEVs provide masses to the sterile
neutrinos. The reason why two B$-$L Higgs fields are needed will be clarified soon.
In a supersymmetric theory, two Higgs fields are anyway required for the anomaly cancellation.
Here we adopt a flavor basis for $N_I$, but the mixing between $N_1$ and $N_i$ is suppressed
in the models considered below.
In the split flavor mechanism, we will introduce a flavor symmetry, under which only the fields in the seesaw
sector are charged, and the SM fields are assumed to be neutral.
The role of the flavor symmetry is to suppress both the mass and mixings of $N_1$ to satisfy the X-ray
bound (\ref{x-ray}), and the key is to assign a flavor charge on one or more of
the B$-$L Higgs fields.
As reference values we take $M_1 \approx 1-10$\,keV and $M_{i} \approx \GEV{14-15}$,
but it is straightforward to further impose a usual FN flavor symmetry, e.g., in order to make
$N_2$ much lighter than $N_3$.

\subsection{Non-supersymmetric case}

We adopt a $Z_4$ flavor symmetry under which only $\Phi^\prime$ and $N_1$ are charged while
the others are singlet:
\vskip 0.5cm
\begin{center}
\begin{tabular}{|c||c|c|c|c|c|c|}
\hline
                    & $\quad\Phi\quad$   & $\quad \Phi^\prime \quad$ & $\quad N_1 \quad$ & $\quad N_i \quad$
                    & $\quad L_\alpha \quad$  &  $\quad H \quad$ \\
\hline
\,\,U$(1)_{{\rm B}-{\rm L}}$\,  &   $2$     &   $-2n$     & $-1$  & $-1$   & $-1$ &  0  \\
\hline
$Z_4$               &   0    &   $-1$     & $1$  &  0    & 0 & 0  \\
\hline
\end{tabular}
\end{center}
\vskip 0.3cm
with $n$ being a positive integer, and $i=2,3$.
Then the seesaw sector is described by
\bea
-\Delta {\cal L} =
\frac{1}{2}\kappa_i \Phi \bar N^c_i N_i + \lambda_i \bar N_i L H
+ \frac{1}{2} \kappa_1 \frac{(\Phi^{2n-1}\Phi^{\prime 2})^*}{\Lambda^{2n}} \bar N^c_1 N_1
+ \tilde \lambda \frac{(\Phi^n\Phi^\prime)^*}{\Lambda^{n+1}} \bar N_1 L H
+ {\rm h.c.},
\eea
for a cut-off scale $\Lambda$.
Here $\kappa_1$, $\kappa_i$, $\lambda_i$ and ${\tilde \lambda}$ are numerical coefficients of order unity,
and we have dropped the lepton flavor indices.
Note that the term $\Phi^{n+1}\Phi^\prime \bar N^c_1 N_i$ has been omitted as it can be removed
by redefining $N_I$ without any significant effects on the above interactions.

\begin{figure}[t]
\begin{center}
\hspace{-0.5cm}
\includegraphics[width=7.5cm,clip,angle=0]{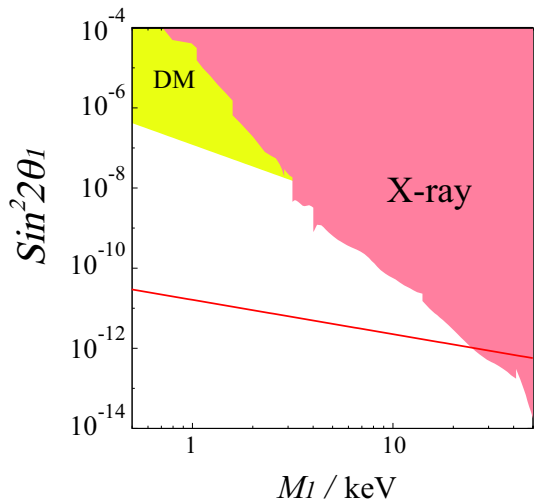}
\vspace{-8mm}
\end{center}
\caption{The mixing angle $\sin^2 2 \theta_1$ in the non-supersymmetric model
with the $Z_4$ flavor symmetry, where we have taken $n=3$ and $\Lambda=M_p$
under the assumption that $\Phi$ and $\Phi^\prime$ have VEVs of a similar size.
The upper-right (pink) and upper-left (yellow) shaded regions are excluded by the
the X-ray observations and the dark matter overproduction via the Dodelson-Widrow mechanism,
respectively.
}
\label{fig:x-ray2}
\end{figure}

The U(1)$_{{\rm B}-{\rm L}}$ gauge symmetry is spontaneously broken when $\Phi$ and $\Phi'$ develop
a non-zero VEV.
Here we assume $\la \Phi \ra \gtrsim \langle \Phi^\prime \rangle$.
As a result, the mass of the two heavy right-handed neutrinos is set by $M=\langle \Phi \rangle$,
and the light neutrino masses are nicely explained by the seesaw mechanism.
The above neutrino interactions lead to the mass and mixing of the $N_1$ as
\bea
M_1 &\approx&
\lrfp{M}{\Lambda}{2(n-1)}  \lrfp{ \langle \Phi^\prime \rangle}{\Lambda}{2} M,\\
\lambda_{1 \alpha} &\approx& \lrfp{M}{\Lambda}{n}  \frac{ \langle \Phi^\prime \rangle}{\Lambda},
\eea
implying
\bea
\epsilon &\approx&  \frac{M}{\Lambda}.
\eea
Therefore the suppression of $\epsilon$ is achieved for $M\ll \Lambda$, and consequently
the active-sterile neutrino mixing is estimated to be
\bea
\theta^2_1 &\approx&
10^{-5}
\lrfp{M}{\Lambda}{2}
\left(\frac{m_{\rm seesaw}}{0.1\,{\rm eV}}\right)
\left(\frac{M_1}{10\,{\rm keV}}\right)^{-1},\non\\
&\simeq&
2\times 10^{-12}
\left(\frac{m_{\rm seesaw}}{0.1{\rm eV}}\right)
\left(\frac{M_1}{10{\rm keV}}\right)^{-1}
\left(\frac{M}{10^{15}{\rm GeV}}\right)^2,
\eea
where we have set $\Lambda$ to be the Planck scale, $M_p \simeq 2.4 \times \GEV{18}$,
in the second equality.
Note that the mixing angle depends on $n$ only through $M_1$.
For instance, in the case of $n=3$, $M_1$ is around 10~keV when both $\Phi$ and $\Phi^\prime$
have a VEV around $10^{15}$~GeV.
Fig.~\ref{fig:x-ray2} shows the property of $N_1$ for the case with $n=3$, assuming that
$\Phi$ and $\Phi^\prime$ have VEVs of a similar size.
Also, $M_1\sim10$\,keV can be realized for $n=1$ or 2 if $\langle \Phi^\prime \rangle$ is at an intermediate
scale, which is possible because  there is no dynamical reason
to relate $\la \Phi \ra$ to $ \la \Phi' \ra$ in contrast to supersymmetric cases.

It is possible to consider a general discrete symmetry $Z_k$ under which only $\Phi^\prime$ and $N_1$
are charged.
A proper $Z_k$ charge assignment makes $N_1$ have a small Yukawa coupling induced from
the term $(\Phi^a \Phi^{\prime b})^* \bar N_1 L H$ after B$-$L breaking.
Here $\Phi^\prime$ carries a B$-$L charge equal to $-2a/b$ for coprime positive integers $a$ and $b$.
Then it is obvious that $M_1$ always receives contribution from $\Phi(\Phi^a \Phi^{\prime b})^2\bar N^c_1 N_1$.
If it is the dominant contribution, one obtains $\epsilon \sim 1$ as in the simple FN model, and thus
the longevity problem is not solved.
This holds also when one uses a global U$(1)$ instead of $Z_k$.
We note that a suppression of $\epsilon$ can be achieved by taking a $Z_k$ charge assignment such
that $N_1$ gets a mass dominantly either from $(\Phi^{2a-1}\Phi^{\prime 2b})^* \bar N^c_1 N_1$
or from $\Phi (\Phi^a\Phi^{\prime b})^* \bar N^c_1 N_1$.

\subsection{Supersymmetric case}

The seesaw mechanism can be embedded into a supersymmetric framework.
For the anomaly cancellation, $\Phi$ and $\Phi^\prime$ must be vector-like under U$(1)_{{\rm B}-{\rm L}}$.
Interestingly enough, it is then possible to suppress $M_1$ as well as the active-sterile neutrino mixing
by both supersymmetry (SUSY) breaking effects and a flavor symmetry.
We will also show that a discrete R-symmetry can do the job.

\subsubsection{Discrete flavor symmetry}
Let us first consider a $Z_k$ flavor symmetry with $k \geq 3$, under which only $\Phi^\prime$ and $N_1$
transform non-trivially and the others are neutral:
\vskip 0.5cm
\begin{center}
\begin{tabular}{|c||c|c|c|c|c|c|}
\hline
                    & $\quad\Phi\quad$   & $\quad \Phi^\prime \quad$ & $\quad N_1 \quad$ & $\quad N_i \quad$
                    & $\quad L_\alpha \quad$ & $\quad H_u \quad$ \\
\hline
\,\,U$(1)_{{\rm B}-{\rm L}}$\,  &   $-2$                  &   $2$       & $1$  & $1$   & $-1$ & 0     \\
\hline
$Z_k$               &   0                     &   1  &1 &  0   & 0 & 0     \\
\hline
\end{tabular}
\end{center}
\vskip 0.3cm
with $H_u$ being the up-type Higgs doublet superfield.
Such discrete symmetry acting on one of the B$-$L Higgs fields was  considered in the B$-$L Higgs
inflation models~\cite{Nakayama:2012dw}.
Note that $N_I$, $\Phi$ and $\Phi^\prime$ are left-chiral superfields, and in particular,
the fermionic component of $N_I$ is the left-handed anti-neutrino.
That is why the B$-$L charge assignment on these fields is different from the non-supersymmetric case.

With the above charge assignment, the relevant terms in the K\"ahler and super-potentials
of the seesaw sector are given by
\bea
\Delta K &=&  \frac{\Phi^{\prime *}}{\Lambda} N_1 N_i
+ \frac{1}{2} \frac{(\Phi\Phi^{\prime2})^*}{\Lambda^3} N_1 N_1 + {\rm h.c.},
\nonumber \\
\Delta W &=& \frac{1}{2}  \Phi N_i N_i + N_i L H_u
+  \frac{(\Phi\Phi^\prime)^{k-1}}{\Lambda^{2k-2}} N_1 L H_u
+ \frac{1}{2} \frac{\Phi (\Phi\Phi^\prime)^{k-2}}{\Lambda^{2k-4}} N_1 N_1,
\eea
where we have omitted coupling constants of order unity.\footnote{
Instead of the discrete symmetry, one can take a global U$(1)$ symmetry under which $\Phi^\prime$ and
$N_1$ have the same charge and the other fields are neutral.
Then the terms in $\Delta K$ are still allowed while the last two terms in $\Delta W$ are forbidden.
The Nambu-Goldstone boson associated with U$(1)$ may contribute to dark
radiation~\cite{Nakayama:2010vs,Weinberg:2013kea}.
}
Though we have not considered  here, one may impose a U$(1)_R$ symmetry under the assumption that
it is broken by a small constant term in the superpotential, i.e. by the gravitino mass $m_{3/2}$.
As we shall see shortly, in such case,  both of the terms in $\Delta K$ can be further suppressed by $m_{3/2}$
if the superpotential is to possess the term $\Phi N_i N_i$.
Note here that the gravitino mass represents the explicit U(1)$_R$ breaking by two units.

To examine the property of sterile neutrino dark matter, it is convenient to integrate out
the U$(1)_{{\rm B}-{\rm L}}$ sector.
The U$(1)_{{\rm B}-{\rm L}}$ is broken along the $D$-flat direction $|\Phi|^2=|\Phi^\prime|^2=M^2$,
which is stabilized by higher dimension operators, or by a radiative potential induced by
the $\lambda_i$ interaction.
For $M$ much larger than the gravitino mass $m_{3/2}$, the effective theory of neutrinos is written as
\bea
\Delta W_{\rm eff} &=&  \frac{1}{2} \kappa_i M N_i N_i
+ \lambda_i N_i L H_u
+\, \frac{1}{2}
M_1 N_1N_1
+ \lambda_{1 \alpha} N_1 L H_u,
\eea
at energy scales around and below $M$, where the sterile neutrino $N_1$ obtains
\bea
\label{susyM1}
M_1 &=& \frac{m_{3/2} M^3}{\Lambda^3} + \frac{M^{2k-3}}{\Lambda^{2k-4}},\\
\label{susylam}
\lambda_{1 \alpha} &=&  \frac{m_{3/2}}{\Lambda} + \frac{M^{2k-2}}{\Lambda^{2k-2}}
\eea
omitting numerical coefficients of order unity.
Here the terms proportional to $m_{3/2}$ arise from $\Delta K$ after redefining $N_i$ to remove
mixing terms $N_1N_i$ in the effective superpotential.
In contrast to the non-supersymmetric case, there are two important effects here.
One is the holomorphic nature of the superpotential, and the other is the SUSY breaking effects represented
by the gravitino mass.

\begin{figure}[t]
\begin{center}
\begin{minipage}{16.4cm}
\centerline{
{\hspace*{-0.5cm}\epsfig{figure=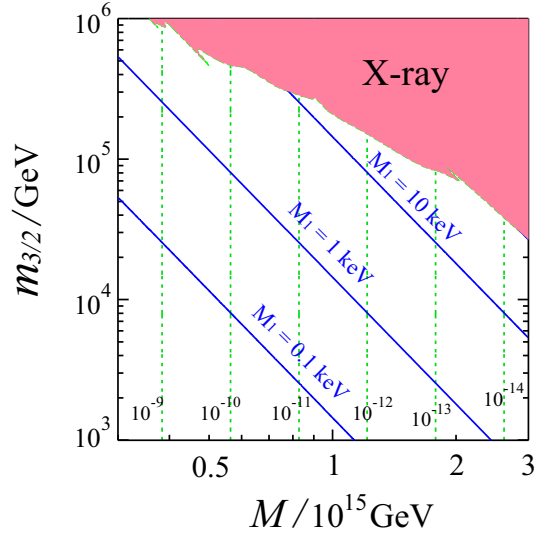,angle=0,width=7.2cm}}
}
\vskip -0.5cm
\caption{
Contours of the sterile neutrino mass $M_1$ (solid (blue)) and the mixing angle $\theta_1^2$
(dashed (green)) in the $M$-$m_{3/2}$ plane for the case of the discrete $Z_k$ with $k\geq 6$.
The upper-right (pink) shaded region is excluded by the X-ray observations. Here we have fixed
the cut-off scale as $\Lambda = M_p$.
}
\label{susyZk}
\end{minipage}
\end{center}
\end{figure}

Depending on the values of $M$, $\Lambda$, $m_{3/2}$ and $k$, there are various possibilities.
To simplify our analysis, let us focus on the case of the reference values, $M \sim \GEV{15}$
and $\Lambda = M_p$.
Then $M_1\sim 10$\,keV is realized for $m_{3/2} \lesssim {\cal O}(100)$\,TeV and $k\geq 5$,\footnote{
This may provide a motivation to consider SUSY around $100$\,TeV,  which is consistent with the recent
discovery of the SM-like Higgs boson of mass $\sim 126$\,GeV.
If the SUSY breaking was much higher, the sterile neutrino could not be dark matter because of its too short
lifetime.
Note that the decay rate is proportional to $M_1^5$.
}
for which the neutrino Yukawa coupling $\lambda_{1\alpha}$ receives the dominant contribution from
the SUSY breaking effect, i.e., from the first term in Eq.~(\ref{susylam}).
Note also that $M_1$ is determined entirely by the SUSY breaking effect for $k \geq 6$.
In the following we consider $m_{3/2} \sim 100$\,TeV and $k \geq 6$.
The $\epsilon$ parameter and active-sterile neutrino mixing angle then read
\bea
\epsilon &\approx& 4 \times 10^{-4} \lrfp{m_{3/2}}{100{\rm TeV}}{\frac{5}{6}}
\lrfp{ M_1}{10{\rm keV}}{-\frac{1}{3}},
\\
\theta_1^2 &=& \epsilon^2 \frac{m_{\rm seesaw}}{M_1} \,\approx\,
10^{-12}
\lrf{m_{\rm seesaw}}{0.1 {\rm eV}} \lrfp{m_{3/2}}{100{\rm TeV}}{\frac{5}{3}}
\lrfp{ M_1}{10{\rm keV}}{-\frac{5}{3}}.
\eea
Thus, the observational constraint (\ref{x-ray}) is naturally satisfied if the gravitino
mass is smaller than or comparable to $100$\,TeV.
In particular, the predicted X-ray flux is just below the observational bound for $m_{3/2} \sim 100$\,TeV.
See Fig.~\ref{susyZk}, where the contours of $M_1$ and $\theta_1^2$ are shown in the $(M, m_{3/2})$ plane.
On the other hand, the squarks and sleptons acquire soft SUSY breaking masses in
the range between about $m_{3/2}/8\pi^2$ and $m_{3/2}$, depending on mediation mechanism.
It is interesting to note that the gravitino mass around $100$\,TeV leads to TeV to
sub-PeV scale SUSY, which can accommodate a SM-like Higgs boson at 126\,GeV within
the minimal supersymmetric SM (MSSM).

Lastly we comment on the case with an approximate global U(1)$_R$ broken by a constant superpotential
term.
The neutrino interactions are then further constrained.
For instance, let us consider the case where $N_I$ and $L_\alpha$ have the same R charge equal
to one while $\Phi$, $\Phi^\prime$ and $H_u$ are neutral.
Then both the terms in $\Delta K$ are further suppressed by the gravitino mass.
As a result, the sterile neutrino mass as well as the neutrino Yukawa couplings are determined by the ratio
of the B$-$L breaking scale to the cut-off scale, and the effect of SUSY breaking is negligibly small.
That is to say, $M_1$ and $\lambda_{1 \alpha}$ receive the dominant contributions from
the second terms in (\ref{susyM1}) and (\ref{susylam}), respectively. For the reference values
$M \sim \GEV{15}$ and $\Lambda  = M_p$, $k$ must be equal to $5$ to realize $M_1 \sim 10$\,keV unless
$m_{3/2}$ is extremely heavy (say, $\GEV{11}$ or heavier).
Then the neutrino Yukawa couplings will become extremely small so that sterile neutrino dark matter becomes
practically stable and the predicted X-ray flux is negligibly small.
Although not pursued here, it may be interesting to
consider the case of $k<5$ where a sterile neutrino dark matter is  much heavier than $10\,$keV
and has a sufficiently small mixing angle.\footnote{See Ref.~\cite{Essig:2013goa}
for the latest X-ray and gamma-ray constrains on such heavy sterile neutrino dark matter.}

\subsubsection{Discrete R symmetry}

Next let us consider a case of discrete R symmetry.
The discrete R symmetry has been extensively studied from various cosmological and phenomenological
aspects.
See e.g. Refs.~\cite{Kurosawa:2001iq,Izawa:1996dv,Dine:2009swa,Dine:2010eb,Harigaya:2013vja,Takahashi:2013cxa}.
Now we show that the split flavor mechanism can be implemented by the discrete R symmetry with
the following charge assignment,
\vskip 0.5cm
\begin{center}
\begin{tabular}{|c||c|c|c|c|c|c|}
\hline
                    & $\quad\Phi\quad$   & $\quad \Phi^\prime \quad$ & $\quad N_1 \quad$ & $\quad N_i \quad$
                    &$\quad L_\alpha \quad$ & $\quad H_u\quad $\\
\hline
\,\,U$(1)_{{\rm B}-{\rm L}}$\,  &   $-2$                  &   $2$       & $1$  & $1$     & $-1$ & 0   \\
\hline
$Z_{kR}$               &   0                    &   $p$     & $q $  &  1      &1&0    \\
\hline
\end{tabular}
\end{center}
\vskip 0.3cm
where $p$ and $q$ are integers mod $k$.
To simplify our analysis, we assume that the cut-off scale for higher dimensional operators is
given by the Planck scale, $M_p$, and the B$-$L breaking scale $M$ is about $\GEV{15}$.
The gravitino mass is assumed to be below PeV scale.

Note that the discrete $Z_{kR}$ symmetry ($k \geq 3)$ is explicitly broken by the constant term
in the superpotential, $\la W \ra \simeq m_{3/2} M^2_p$.
Therefore, the mass $M_1$ and neutrino Yukawa couplings $\lambda_{1\alpha}$ generically receive two
contributions; one is invariant under $Z_{kR}$, and the other is not invariant and is proportional
to the gravitino mass.

The sterile neutrino mass $M_1 \sim 10$\,keV is numerically close to $M^7/M_p^6$ or $m_{3/2} M^3/M_p^3$, and
the mass of this order can be generated  if one or more of the following operators are allowed:
\bea
\Delta K &=& \frac{(\Phi\Phi^{\prime 2})^*}{M_p^3} N_1 N_1  + {\rm h.c.},
\nonumber \\
\Delta W &=& \frac{\Phi(\Phi\Phi^\prime)^3}{M^6_p} N_1N_1 \,\,\,{\rm or}\,\,\,
m_{3/2} \frac{\Phi^2 \Phi'}{M_p^3} N_1 N_1.
\eea
Similarly, the neutrino Yukawa coupling of the desired magnitude can be induced from the following
operators,
\bea
\Delta K &=&
\frac{\Phi^{\prime *}}{M_p} N_1N_i + {\rm h.c.},
\nonumber \\
\Delta W &=&
\frac{(\Phi\Phi^\prime)^2}{M^4_p} N_1 L H_u \,\,\,{\rm or}\,\,\, \frac{m_{3/2}}{M_p}N_1 L H_u.
\eea
In order for one or more of the above operators to give the dominant contribution
to $M_1$ and $\lambda_{1\alpha}$, the following operators must be forbidden by the discrete R-symmetry:
\bea
\Delta K_{\rm forbidden} &=& \frac{\Phi^{\prime *}}{M_p} N_1 N_1 + {\rm h.c.},
\nonumber \\
\Delta W_{\rm forbidden} &=&  \Phi N_1 N_I + \frac{\Phi^2\Phi^\prime}{M_p^2} N_1 N_I +
\frac{\Phi(\Phi\Phi^\prime)^2}{M^4_p} N_1 N_1 + N_1LH_u +
\frac{\Phi \Phi'}{M_p^2} N_1 LH_u,
\eea
which puts constraints on $p$ and $q$.

To summarize, we need to find a set of $(k,p,q)$ satisfying
\bea
&& \hspace{-0.5cm} 2p-2q \equiv 0 ~~~{\rm or}~~~3p + 2 q \equiv 2 ~~~{\rm or}~~~p+2 q \equiv 0, \\
&& \hspace{-0.5cm} p-q-1\equiv 0 ~~~{\rm or}~~~2p+q +1\equiv 2 ~~~{\rm or}~~~q+1 \equiv 0,  \\
&& \hspace{-0.5cm} p-2q \not\equiv 0, ~~2q \not\equiv 2, ~~q+1 \not\equiv 2,
~~p+2q \not\equiv 2, ~~p+q+1 \not\equiv 2,
~~2p+2q \not\equiv 2,
\eea
where all the equations are mod $k$.
Some of the solutions of the above conditions are\footnote{
If we forbid a SUSY mass $\Phi \Phi'$ in the superpotential, the solutions with $p = 2$
should be excluded.
}
\beq
(k,p,q) = (5,2,2),\,(5,4,3),\,(7,3,2),\,(7,5,4),\,(7,5,5),\,(7,6,6),\,\cdots.
\eeq
In fact there is no solution for which both $M_1$ and $\lambda_{1\alpha}$ are generated
by the $Z_{kR}$ invariant operators. That is to say, either or both of them should be
generated by the SUSY breaking effect proportional to the gravitino mass.

\begin{figure}[t]
\begin{center}
\begin{minipage}{16.4cm}
\centerline{
{\hspace*{-0.5cm}\epsfig{figure=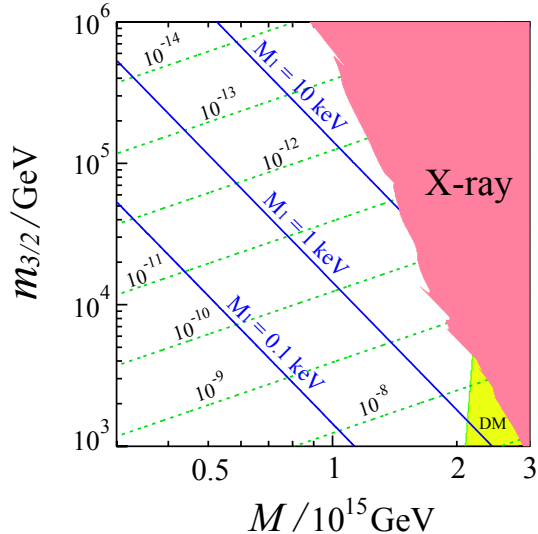,angle=0,width=7.2cm}}
}
\vskip -5mm
\caption{
Contours of the sterile neutrino mass $M_1$ (solid (blue)) and the mixing angle $\theta_1^2$
(dashed (green)) in the $M$-$m_{3/2}$ plane for the case of the discrete R symmetry.
The upper-right (pink) and lower-right (yellow) shaded
region  are excluded by the X-ray observations and the dark matter overproduction via
the Dodelson-Widrow mechanism, respectively.
}
\label{susyZ5R}
\end{minipage}
\end{center}
\end{figure}

Let us focus on the case of $(k,p,q) = (5,4,3)$.
Then the relevant terms in the superpotential are given by
\bea
\Delta W &=& \frac{1}{2} \Phi N_i N_i +  N_i L H_u +
\frac{1}{2} m_{3/2} \frac{ \Phi^2 \Phi^\prime}{M_p^3} N_1 N_1
+ \frac{(\Phi\Phi^\prime)^2}{M^4_p} N_1 L H_u,
\eea
where we have dropped numerical coefficients of order unity.
The other interactions in the K\"ahler and super-potentials are either forbidden or irrelevant
for the following discussion.
The mass and neutrino Yukawa couplings for $N_1$ are given by
\bea
M_1 & \approx & 10{\rm\,keV} \lrf{m_{3/2}}{100{\rm TeV}} \lrfp{M}{\GEV{15}}{3} ,\\
\lambda_{1\alpha} & \approx& 10^{-14} \lrfp{M}{\GEV{15}}{4},
\eea
from which one finds
\bea
\epsilon &\simeq& 3 \times 10^{-4} \lrfp{m_{3/2}}{100{\rm TeV}}{-\frac{1}{2}} \lrfp{M}{\GEV{15}}{3},
\eea
using the D-flat condition, $\la \Phi \ra = \la \Phi' \ra = M$.
Therefore the mass $M_1$ is close to $10$\,keV and $\epsilon \sim 10^{-3}$ for the reference values $M = \GEV{15}$
and $\Lambda = M_p$.
Finally, the mixing angle reads
\bea
\theta_1^2 &\approx& 2\times10^{-12} \lrf{m_{\rm seesaw}}{0.1{\rm\,eV}}
\lrf{M_1}{10{\rm keV}} \lrfp{m_{3/2}}{100 {\rm TeV}}{-3}.
\eea
We show the contours
of $M_1$ and the mixing angle $\theta_1^2$ are shown in the $M$-$m_{3/2}$ plane in Fig.~\ref{susyZ5R}.
It is interesting to note that $m_{3/2} \sim 100$\,TeV and $M \sim \GEV{15}$ lead to
the sterile neutrino mass $M_1 \sim 10$\,keV with the predicted X-ray line flux just
below the current bound.

\section{Cosmological aspects}
We have so far focused on the mass and mixing angles of the sterile neutrinos.
In order for the lightest sterile neutrino $N_1$ to account for  the observed dark matter,
a right amount of $N_1$ must be produced in the early Universe.
The density parameter of dark matter is related to the number to entropy ratio $n_{N_1}/s$ as
\bea
\Omega_{\rm DM} h^2 &\simeq& 0.14 \lrf{M_1}{10{\rm\,keV}} \lrf{n_{N_1}/s}{5 \times 10^{-5}},
\eea
where $h$ is the dimensionless Hubble parameter in the units of $100\,{\rm km}\,s^{-1} {\rm Mpc}^{-1}$,
and $n_{N_1}$ and $s$ are the number density of $N_1$ and the entropy density, respectively.
The latest observations give $\Omega_{\rm DM}h^2 \simeq 0.1199 \pm 0.0027$~\cite{Ade:2013zuv}.

The thermal production known as the Dodelson-Widrow mechanism~\cite{Dodelson:1993je}
is in tension with the X-ray bound for $M_1 \gtrsim 10$\,keV, as can be seen from
Fig.~\ref{fig:x-ray}.
Therefore we need another production mechanism.
One possibility is that the $N_1$ is  produced via the s-channel exchange of the B$-$L gauge
boson~\cite{Kusenko:2010ik}.
The number to entropy ratio of the sterile neutrino produced by this mechanism is roughly estimated as
\bea
\frac{n_{N_1}}{s}
&\sim & 10^{-4} \lrfp{g_*}{100}{\frac{3}{2}} \lrfp{M}{\GEV{15}}{-4} \lrfp{T_R}{5 \times \GEV{13}}{3},
\label{thermal}
\eea
where  $g_*$ counts the relativistic degrees of freedom at the reheating, and $T_R$ denotes
the reheating temperature.
The numerical solution of the Boltzmann equation gives a consistent result~\cite{Khalil:2008kp}.
The assumption here is that the B$-$L symmetry is spontaneously broken during and after inflation.
This production mechanism works both for supersymmetric and non-supersymmetric cases.
Also, a right amount of the baryon asymmetry can be created via thermal leptogenesis due to the two heavy
right-handed neutrinos $N_2$ and $N_3$ for such high reheating temperature~\cite{Endoh:2002wm,Raidal:2002xf}.\footnote{
Thermal leptogenesis in the neutrino mass anarchy hypothesis was studied in Ref.~\cite{Jeong:2012zj}.
}

On the other hand, if the B$-$L symmetry is restored during or after inflation, the sterile neutrinos will
be in thermal equilibrium through the U$(1)_{{\rm B}-{\rm L}}$ gauge interactions.
The thermal abundance is given by
\beq
\frac{n_{N_1}^{\rm (eq)}}{s} \;\simeq\;  2 \times 10^{-3} \lrfp{g_*}{100}{-1}.
\eeq
So, if there is an entropy dilution of the order of a few tens, the right amount of $N_1$ can be generated.
In the non-supersymmetric case, such entropy dilution can be easily realized by the B$-$L Higgs dynamics. Suppose
that the mass of the B$-$L Higgs is slightly smaller than the  B$-$L breaking scale. Then it remains
trapped at the origin due to the thermal mass induced by the B$-$L gauge boson loop, dominating the
Universe for a while. This is a mini-thermal inflation.\footnote{
See Ref.~\cite{Yamamoto:1985rd} for the usual thermal inflation.
The entropy production due to the bubble formation was discussed in Ref.~\cite{Kusenko:2010ik}.
}
When the plasma temperature becomes lower than the mass, the B$-$L Higgs develops a large VEV, and
the subsequent decays of the B$-$L Higgs produce the entropy.
Also, thermal and/or non-thermal leptogenesis works successfully in this case.
Since we have imposed a discrete symmetry on the B$-$L Higgs, domain walls are generally produced.
The domain walls will annihilate if we add a small breaking of the discrete symmetry. Interestingly, gravitational
waves~\cite{GW} are likely produced during the violent annihilation
processes of the domain walls~\cite{Gleiser:1998na,Takahashi:2008mu,Dine:2010eb,Hiramatsu:2010yz,
Kawasaki:2011vv,Hiramatsu:2013qaa},
which may be within the reach of the future and planned
gravitational wave experiments. After the domain wall annihilation, we are left with the cosmic strings whose tension is 
consistent with the CMB observation~\cite{Ade:2013xla} for $M \lesssim  \oten{15}$\,GeV.

In a supersymmetric case, on the other hand, the stabilization of the B$-$L Higgs is slightly more involved.
To be concrete, let us consider the model based on the discrete R symmetry and adopt
$(k,p,q) = (5,4,3)$ in the following.
The $D$-flat direction composed of $\Phi$ and $\Phi'$ can be stabilized by the balance between
non-renormalizable superpotential term $\phi^6/M_p^3$ and SUSY breaking effect
(negative soft mass squared at the origin, or the $A$-term associated with the superpotential term):
\beq
V \;=\; -m_{\phi}^2 |\phi|^2 - \left( m_{3/2} \frac{\phi^6}{M_p^3} +{\rm h.c.} \right)+
 \frac{|\phi|^{10}}{M_p^6},
\eeq
where $\phi^2 \equiv \Phi\Phi'$ parameterizes the $D$-flat direction, $m_\phi^2$ represents the soft mass for
the $D$-flat direction, and we have dropped numerical coefficients of order unity.
The B$-$L Higgs is then stabilized at
\beq
M  = \la \phi \ra \sim \GEV{15} \lrfp{m_{3/2}}{100\,{\rm TeV}}{\frac{1}{4}}.
\eeq
If the U(1)$_{\rm B-L}$ symmetry is restored during or after inflation, thermal inflation generically takes place
because $\phi$ has a relatively flat potential. Then the entropy dilution factor tends to be large, and any
pre-existing $N_1$ will be diluted away. The subsequent domain walls can be erased if we introduce a
breaking of the discrete symmetry.\footnote{
In the case of the discrete R symmetry, the constant term in the superpotential provides such breaking terms.
Unfortunately, however, its size is too small to make domain walls to annihilate before dominating the Universe.
}

In the supersymmetric case, the lightest supersymmetric particle (LSP) in the MSSM
contributes to the dark matter abundance. Even though the R-parity is broken
in the case of the discrete $Z_{5R}$ symmetry, the MSSM-LSP is stable due to the residual $Z_{2\,{\rm B-L}}$ since
U(1)$_{\rm B-L}$ is spontaneously broken only by $\Phi$ and $\Phi^\prime$ with the B$-$L charge two.
In order for the lightest sterile neutrino $N_1$ to be the dominant component of dark matter,
the MSSM-LSP abundance must be suppressed.
If the reheating temperature is as high as $\oten{13}$\,GeV, the Universe becomes {\it gravitino-rich}, and
the  MSSM-LSPs tend to be  overproduced by the gravitino decay~\cite{Jeong:2012en}. The MSSM-LSP abundance can be
suppressed if  it is a Wino-like or Higgsino-like neutralino of mass ${\cal O}(100)$\,GeV and the gravitino mass
is of order PeV.
Since they comprise only a fraction of the total dark matter, the constraints from
indirect dark matter searches are relaxed. It would be interesting if we could see the indirect dark matter
signatures for both the sterile neutrino and the  Wino-like or Higgsino-like neutralino.
On the other hand, if the gravitino mass is of ${\cal O}(100)$\,TeV, the MSSM-LSPs are overproduced
by the gravitino decay.
It is actually possible to make the MSSM-LSP unstable. Let us consider the case of the discrete
R symmetry with $(k,p,q) = (5,4,3)$.  Then, this can be achieved by introducing
another vector-like pair of the B$-$L Higgs $\varphi (1,-1)$ and ${\bar \varphi} (-1,1)$ where the B$-$L
and R-charges are shown in the parenthesis, respectively. If $\varphi$ and ${\bar \varphi}$ have
a nonzero VEV, say, of $\oten{6}$\,GeV, the trilinear
R-parity violating operators are allowed, and the MSSM-LSP decays before the big bang nucleosynthesis.
The constraints from the proton decay can be safely satisfied~\cite{Hinchliffe:1992ad}.
Alternatively, if there is another production mechanism of the sterile
neutrino dark matter which works at a temperature below $\GEV{9}$, the Universe is not gravitino-rich,
and we can avoid the overproduction of the MSSM-LSPs from the gravitino decay.

\section{Conclusions}

The sterile neutrino dark matter of mass ${\cal O}(1-10)$\,keV generically decays into an active neutrino
and an X-ray photon,
but the non-observation of the X-ray line requires the sterile neutrino to be more long-lived
than estimated based on the seesaw formula. Specifically, the neutrino Yukawa couplings
$\lambda_{1\alpha}$ must be suppressed by more than two orders of magnitude than naively
estimated for $M_1 = 10$\,keV.
We call this tension as the longevity problem for the sterile neutrino dark matter.
It is worth noting that the longevity problem is not solved by the simple FN model and the split seesaw
mechanism, both of which preserve the seesaw formula.
In this paper we have  quantified the longevity problem and proposed the split flavor mechanism as
a possible solution.
In this mechanism, we have introduced a single flavor symmetry (or discrete R symmetry) under which
one or more of the B$-$L Higgs is charged. As a result, the split mass spectrum for the sterile
neutrinos as well as the longevity required for the lightest sterile neutrino dark matter are realized.
The key is to combine the B$-$L symmetry with the flavor symmetry.
We have provided several examples in which the lightest sterile neutrino of mass is  ${\cal O}(1-10)$ keV and
the predicted X-ray flux is just below the current bound.  Therefore it may possible
to test our models in the future X-ray observations.

\section*{Acknowledgment}
This work was supported by
Grant-in-Aid for Scientific Research (C) (No. 23540283) [KSJ],  Scientific Research on Innovative
Areas (No.24111702 [FT], No. 21111006 [FT] , and No.23104008 [KSJ and FT]), Scientific Research (A)
(No. 22244030 and No.21244033) [FT], and JSPS Grant-in-Aid for Young Scientists (B)
(No. 24740135) [FT], and Inoue Foundation for Science [HI and FT].
This work was also supported by World Premier International Center Initiative
(WPI Program), MEXT, Japan [FT].

\end{document}